\documentclass{article}
\usepackage{preprint}

\usepackage{microtype}
\usepackage{graphicx}
\usepackage{subcaption}
\usepackage{booktabs} 

\usepackage{amsmath,amssymb}

\usepackage[square]{natbib}
\usepackage[utf8]{inputenc}	
\usepackage[T1]{fontenc}	
\usepackage{xcolor}		
\usepackage[colorlinks = true,
            linkcolor = purple,
            urlcolor  = blue,
            citecolor = cyan,
            anchorcolor = black]{hyperref}	
\usepackage{booktabs} 		
\usepackage{nicefrac}		
\usepackage{microtype}		
\usepackage{lineno}		
\usepackage{float}			

\usepackage{lipsum}		

\usepackage{newfloat}
\DeclareFloatingEnvironment[name={Supplementary Figure}]{suppfigure}
\usepackage{sidecap}
\sidecaptionvpos{figure}{c}

\usepackage{titlesec}
\titlespacing\section{0pt}{12pt plus 3pt minus 3pt}{1pt plus 1pt minus 1pt}
\titlespacing\subsection{0pt}{10pt plus 3pt minus 3pt}{1pt plus 1pt minus 1pt}
\titlespacing\subsubsection{0pt}{8pt plus 3pt minus 3pt}{1pt plus 1pt minus 1pt}

\usepackage{tikz,xcolor,hyperref}

\definecolor{lime}{HTML}{A6CE39}

\def\TA{\mathbf A}
\def\TD{\mathbf D}
\def\TM{\mathbf M}

\def\bh{\mathbf h}
\def\TK{\mathbf K}
\def\TX{\mathbf X}

\title{Machine Learning for Airborne Electromagnetic Data Inversion: a Bootstrapped Approach}

\usepackage{authblk}

\author[1\thanks{\tt{ophir@compgeoinc.com}}]{Ophir Greif}
\author[1\thanks{\tt{bas@compgeoinc.com}}]{Bas Peters}
\author[2]{Michael S. McMillan}
\author[1]{Paulina Wozniakowska}
\author[3]{Eldad Haber}

\affil[1]{Computational Geosciences Inc. Vancouver, BC, Canada}
\affil[2]{Invert Geophysics, Svolvaer, Norway}
\affil[3]{University of British Columbia, Vancouver, BC, Canada}

\begin{document}

\maketitle

\begin{abstract}
Aircraft-based surveying to collect airborne electromagnetic data is a key method to image large swaths of the Earth's surface in pursuit of better knowledge of aquifer systems. Despite many years of advancements, 3D inversion still poses challenges in terms of computational requirements, regularization selection, hyperparameter tuning and real-time inversion. We present a new approach for the inversion of airborne electromagnetic data that leverages machine learning to overcome the computational burden of traditional 3D inversion methods, which implicitly includes learned regularization and is applicable in real-time. The method combines 1D inversion results with geostatistical modeling to create tailored training datasets, enabling the development of a specialized neural network that predicts 2D conductivity models from airborne electromagnetic data. This approach requires 3D forward modeling and 1D inversion up front, but no forward modeling during inference. The workflow is applied to the Kaweah Subbasin in California, where it successfully reconstructs conductivity models consistent with real-world data and geological drill hole information. The results highlight the method's capability to deliver fast and accurate subsurface imaging, offering a valuable tool for groundwater exploration and other near-surface applications.
\end{abstract}

\section{Introduction}
With the world's population growing, rainfall patterns changing, and uncertainty around the long-term drawing of water from aquifers, there is an increased emphasis placed on improving the mapping of known aquifers and discovering new groundwater sources. One of the few geophysical methods that has been proven for rapidly surveying large areas in the search for aquifers is the airborne electromagnetic (AEM) method, see, e.g., \cite{viezzoli2010accurate,doi:10.1071/EG10003, MinsleyTDMC, AbrahamAEMwater, BedrosianComparison, McMillan2018, McMillan2019, Kang2022, Knight_2022, Christensen3DHydro} for case studies. However, while helicopters can quickly collect densely sampled recordings of induced electromagnetic responses along flight lines, data inversion is the current bottleneck. AEM inversion in 3D \citep{Zhang3DEM, Haber2007,doi:10.1071/EG10003,doi:10.1190/geo2011-0370.1, McMillan2015,ren2020three} is demanding both in terms of computation and in terms of the time that it takes an expert to select appropriate parameters, such as regularization, discretization, and meshing. 

While 1D \citep{Farquharson1996, Lat1DInvFixedWing,viezzoli2008spatially, MinsleyTDMC,brodie2012transdimensional} and 2D \citep{wolfgram2003approximate, Wilson2006} approximations can speed up computations, corresponding inversion results suffer from well-known limitations originating from 3D variations in the subsurface conductivity \citep{inv1Dvs25D,ellis1998inversion,wolfgram2003approximate,lin2019discussion,deleersnyder2022assessing}. Despite significant drawbacks, 1D inversions have proven helpful for environments with limited 3D variability. 
Nonetheless, relying on 1D inversion for mapping the subsurface can lead to inaccuracies which we aim to avoid in this work.

The main reason for the computational complexity of AEM inversion stems from the need to solve Maxwell's Equations for every source. Since the AEM problem can have tens of thousands to millions of sources, solving the forward problem is a major computational bottleneck. Thus, 
the first goal of this work is to speed up the inversion of AEM data by avoiding the solution of the forward problem during inference. Instead, Maxwell's Equations will be solved in advance and used to train a machine learning system. 
The second goal is to construct a machine-learning approach that can handle AEM data with varying survey parameters such as transmitter heights, i.e., flight heights. Inversion techniques that fall into this family are referred to as {\em likelihood-free methods} \citep{hamilton2018deep,sainsbury2022fast,deng2022openfwi,chung2024paired}, and they are particularly attractive when the forward problem is difficult to solve.

The following sections introduce a novel, multi-branch neural network tailored to the above-stated goals. We also detail how to construct a training data set for a geological scenario with water aquifers. 

\subsection{Related work}
Neural networks have recently been at the forefront of electromagnetic (EM) research, focusing on speeding up EM inversions and making the results less dependent on difficult-to-tune regularization parameters and other hyperparameters. 
Including neural networks in AEM inversion algorithms has been effective in various ways. For instance, \cite{asif2022integrating} solve the inverse problem using gradient-based optimization but with forward and derivative computations replaced with a neural network. While such an approach can be beneficial, it is limited since the results of this approach can never supersede standard inversion methods in terms of their quality.

Another class of algorithms, closer to our work, replaces the inversion algorithm entirely with a neural network so that the forward problem is not solved during inference. The work on AEM inversion by \cite{Wu2022} maps each time-decay curve and receiver elevation measurement to a resistivity profile, i.e., a 1D solution. See \cite{rs12203440, Bing1DNN, Li1DDNN, Noh1DDNN,bang2022deep, doi:10.1190/geo2022-0723.1, Wu2023fastBayesian,10.1093/gji/ggae244} for other 1D approaches that process each decay-curve and flight-height pair independently. \cite{AEMRNN} use a recurrent neural network to predict 2D subsurface models as a spatial sequence from frequency domain AEM data for ore-body geology. See \cite{PUZYREV2021104681} for deep learning applications to 1D inversion for controlled source marine and land EM. \cite{puzyrev2019deep,9733939,groundTEMDNN} present methods for land-based EM in terms of lines or grids of receivers, with an emphasis on spatially limited anomalies. These works operate in an `image-to-image' fashion to learn a `data-to-model' map. The main drawback of these techniques is that they assume fixed and repeatable surveys so that the data always `looks' the same. Such assumptions break when the flight heights, time channels, source loop size or source position relative to the receiver changes. This makes such approaches less robust in practical scenarios. 

Constructing sufficiently large datasets that represent a given geological scenario is crucial to train the network. Synthetic models are commonly used to train neural networks in the absence of true-earth models. The above-cited papers mainly focus on 1D or homogeneous models with few anomalies, which generally have light computational burdens, allowing for constructing a relatively exhaustive collection of conductivity models. Generating 2D or 3D models is more challenging in terms of creating realistic earth structures and corresponding conductivity values. As we show next, our approach alleviates some challenges by combining fast 1D inversions with geostatistical tools for specific geologies. 

\subsection{Contributions}
We summarize our main contributions to the problem of AEM imaging of real geological structures as follows:
\begin{itemize}
    \item A bootstrapped approach that combines the strengths of computationally inexpensive 1D inversions and synthetic geostatistical modeling to generate training models for a specific type of geology.
    \item A neural network design tailored to predict conductivity models in depth from time-domain AEM data with varying acquisition parameters. We predict a slice from a 3D conductivity model along the flight line.
    \item A demonstration of our full `data-to-model' workflow applied to the Kaweah Subbasin in the Central Valley of California, U.S.A, where we only carry out 3D forward modeling during the dataset generation phase and obtain conductivity models within minutes.
\end{itemize}

\begin{figure}
	\includegraphics[width=0.8\columnwidth]{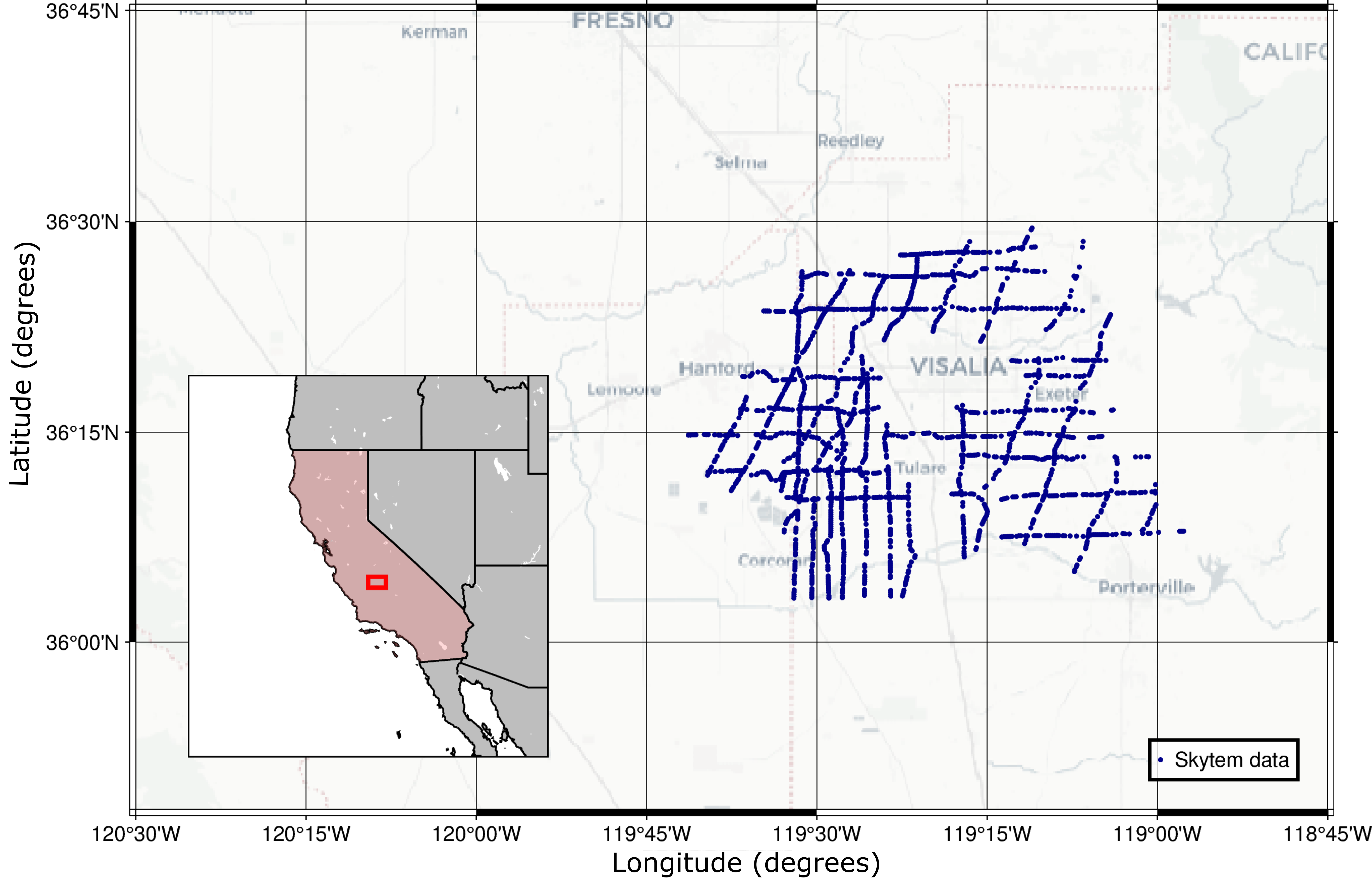}
	\caption{Map of the Kaweah Subbasin in California along with the SkyTEM flight line locations shown in plan view.}
	\label{fig:cali_map}
\end{figure}

After introducing the details of the problem of interest, we lay out the motivation for our network design and its training procedure. Next, we introduce the synergy between 1D AEM inversions and geostatistical geology modeling, then show some results on synthetic validation models. Finally, we show our AI predictions on a field AEM dataset, a SkyTEM \citep{sorense2004skytem} survey from the Kaweah Subbasin in the Central Valley, California, USA (Fig. \ref{fig:cali_map}), and we compare our results to 1D AEM inversions. Geological constraints from nearby boreholes also help evaluate the various models.

\section{Problem setup}
This study aims to develop a machine learning-based alternative to computationally intensive 3D AEM inversions. Notably, we strive to eliminate solving the forward problem during inference. Instead, our approach requires solving a collection of forward problems during the training set generation.

AEM data acquisition yields along the flight line a time series per measurement location (station) per spatial component, and the height of the transmitter and receiver above the ground surface. In this study, we focus only on z-component $d{\bf B}_z/dt$ data, but the method applies to x and y components of the time derivatives of the magnetic field as well as the $\mathbf{B}$-field measurements. For one flight line, the data, $\TD$ is typically a tensor of $c \times n_t \times n_s$, where $n_t$ corresponds to the number of time gates and $n_s$ corresponds to the number of stations and the so-called channel, $c$ represent different types of data\footnote{note that the channels in $c$ represent different data types and not different time channels}. If three-component data is collected, then each one is a different channel in the $c$ dimension. In some surveys, different types of transmitter waveforms are used simultaneously (e.g. low and high-moment). In this case, each channel in $c$ represents the data corresponding to a different waveform. 

In terms of the flight heights, they are collected in the vector $\bh \in \mathbb{R}^{n_s}$. Then, the data are resampled to a uniform $10$m grid, providing the neural networks with a uniform input. Using the data $\TD$ tensor and the flight height $\bh$, we estimate a 2D conductivity model, $\TM$, along the flight line. The conductivity model, $\TM$, is an $n_z \times n_x$ matrix, where $n_z$ and $n_x$ are the number of pixels in the vertical and lateral direction, respectively.

In the following sections, we detail our design for a neural network that achieves this goal and show how we construct realistic training data so that our approach generalizes to field data. 

\section{Designing and training a neural network for AEM inversion}

This section discusses how to create a neural network that transforms time-domain AEM data into depth-domain conductivity models. It is natural to think about this problem as an 'image to image' transform and, therefore, attempt to use convolutional ResNets \citep{He2016} or U-nets \citep{Ronneberger2015} to achieve this task. 
However, several factors prevent this relatively straightforward approach from succeeding. 

First, in many geophysical data surveys, there is variability in different aspects: the source and receiver positioning, flight heights above ground, the source waveform, and receiver time-channels or frequencies. These parameters influence the acquired measurements, resulting in data that appears quite different from one survey to the next. This, in turn, prevents the network from learning how to map the data patterns into a model. Therefore, we need a network design that incorporates survey information so that it can learn to take these parameters into account. 

Secondly, it is well known that convolutional residual networks and U-nets are excellent general-purpose computer vision tools suitable for deblurring, denoising, boundary detection and segmentation, i.e., relatively local image processing tasks. Convolutional networks face more challenges when data needs to be `transported' over spatially large areas \citep{NEURIPS2020_4b21cf96, zakariaei2024advection}, for instance, when late-time data contributes to near-surface conductivities. Therefore, we introduce a network that still uses convolutions but is more adaptable to this situation.

Our network is designed to work with SkyTEM data which uses both high and low moment data.
Using the notation in Table \ref{tab:network_notation} and from the previous section, the aim is to learn to map $\{\TD, \bh\} \rightarrow \TM$. The network takes as input high-moment and low-moment AEM data $\TD$ (two matrices) and the transmitter heights $\bh$ (a vector). We use 16 time channels for both the high-moment and low-moment data in this study with time channels between 43 ms and 3145 ms for the high-moment and 59 ms and 560 ms for the low-moment data. Future networks will aim to use all available time channels as we are missing valuable information by limiting the network to only 16 time channels for each waveform.
\begin{table}
    \centering
    \begin{tabular}{cl}
    \hline
        \textbf{Notation} & \\
        \hline
        $\TX_i$ & network state at layer $i$\\
        $\TK_i$ & block convolution matrix for layer $i$\\
        $\sigma(\cdot)$ & SiLU nonlinear activation\\
        $N(\cdot)$ & instance normalization function\\
        $P_t(\cdot)$ & average pooling in time\\
        $\operatorname{mean}_t$ & reduces time-dimension to its mean\\
        $\TA$ & matrix\\
        $t$ & (artificial) time-step for ResNet\\
        \hline
    \end{tabular}
    \caption{Notation of building blocks for the neural networks. Our convolutions use $5 \times 5$ kernels.}
    \label{tab:network_notation}
\end{table}

To reconcile the different nature of the datatypes, consider the trainable flight-height embedding network $E (\bh,\theta_E) : \mathbb{R}^{n_s} \rightarrow \mathbb{R}^{n_{c1} \times n_t \times n_s}$ that maps the flight heights for $n_s$ stations (measurement locations) into a tensor of $n_{c1}$ network channels $\times$ $n_t$ measurement times gates $\times$ $n_s$ stations. The symbol $\theta_E$ refers to all network weights and biases. This operation enables the addition or concatenation of the now embedded flight heights with the AEM data of size $2 \times n_t \times n_s$, where the two input channels are the low-moment and high-moment data (see Table \ref{network_design_E} for details).

\begin{table}
\caption{Network design for the embedding of flight heights, $E (\bh,\theta_E)$.}
\label{network_design_E}
\begin{center}
\begin{tabular}{rrrl}
\hline 
\multicolumn{1}{c}{\bf Layer} &\multicolumn{1}{c}{\bf Channels} &\multicolumn{1}{c}{\bf Feature size} &\multicolumn{1}{c}{\bf type}\\
\hline
\text{input} & 1     & $128$ & $\bh$\\
1            & 64    & $128$ & $\TX_{1} = \sigma(\TK_1 \bh)$\\
2            & 64    & $128$ & $\TX_{2} = \TK_2 \TX_1$\\
3            & 64    & $16 \times 128$ & $\TD_1$: replicate $\TX_2$ along time-axis\\
\hline
\end{tabular}
\end{center}
\end{table}

The second network converts the time-dependent features into depth-dependent ones while ingesting and merging the already embedded flight heights and AEM data. Denote this network as $F (\TD, E (\bh,\theta_E), \theta_F)$. The time-to-depth conversion is not a simple change of feature size from $n_t$ to $n_z$. To enable the entire time series to influence the full depth range easily, we gradually reduce the time dimension to one, while increasing the number of network channels. This approach is reminiscent of using operations like the Haar transform to exchange time or space for channels in order to increase the receptive field faster \citep{lensink2019fully}. This second network then reassigns the channel dimension over to depth and then outputs the result in terms of a depth dimension and a horizontal dimension. In summary, the second network maps $F (\TD, E (\bh,\theta_E), \theta_F) : \mathbb{R}^{n_{c1} \times n_t \times n_s} \times \mathbb{R}^{n_{c1} \times n_t \times n_s} \rightarrow \mathbb{R}^{n_{c2} \times n_z \times n_s}$. Table \ref{network_design_F} contains the network details.

\begin{figure*}
	\includegraphics[ width=\linewidth]{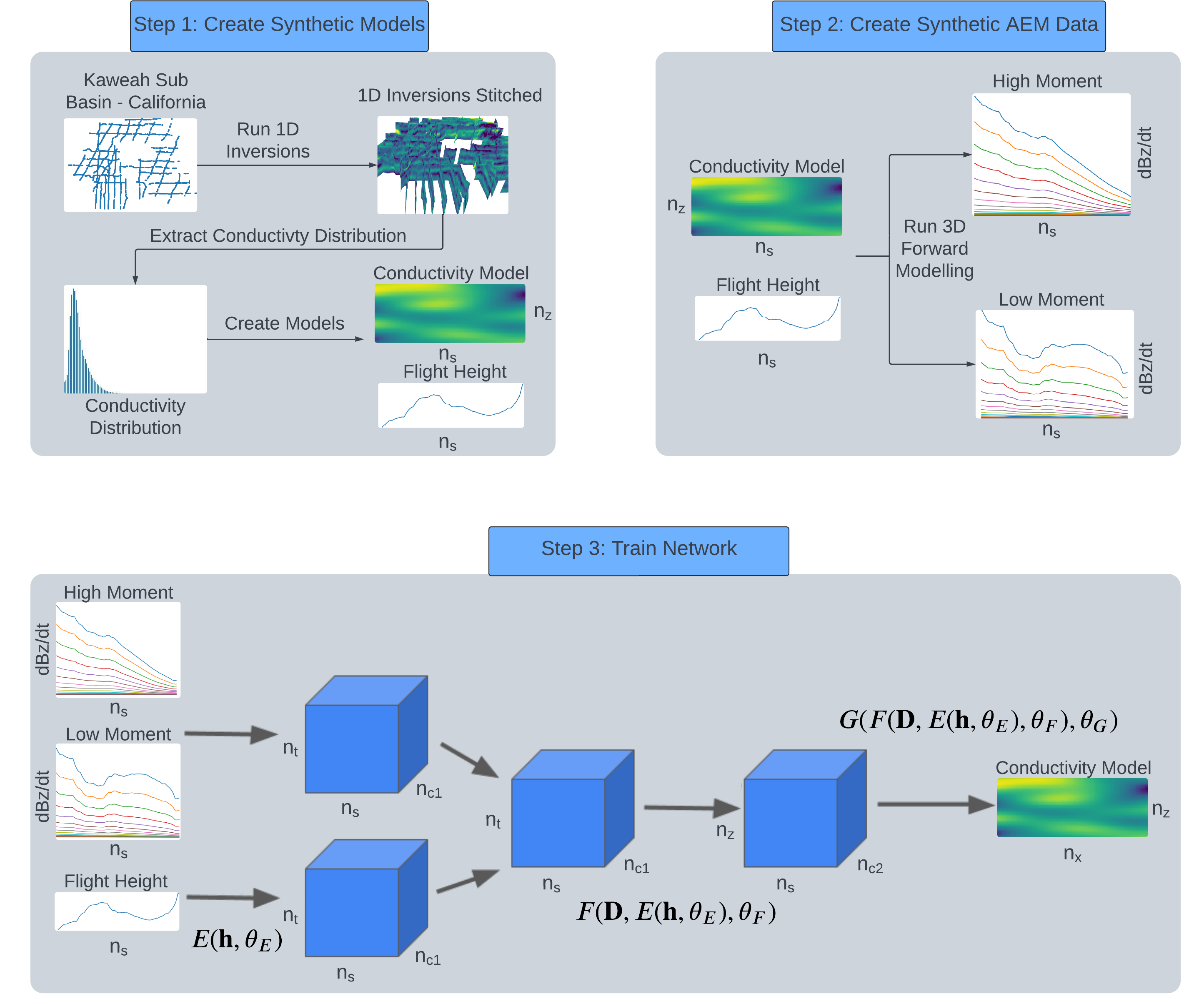}
	\caption{Data pipeline, $n_{t}$ = number of time channels, $n_{s}$ = number of stations, $n_{ci}$ = number of neural network channels in the $\mathit{i}$th network, $n_{x}$ = number of cells in the x-direction, $n_{z}$ = number of cells in the z-direction. $\mathit{E}$ = flight height embedding network, $\mathbf{h}$ = flight height vector, $\theta_E$ = weights and biases for network $\mathit{E}$, $\mathit{F}$ = network to convert from time to depth, $\mathbf{D}$ = AEM data matrices, $\theta_F$ = weights and biases for network $\mathit{F}$, $\mathit{G}$ = network to resample to final conductivity model dimensions, $\theta_G$ = weights and biases for network $\mathit{G}$.}
	\label{fig:flowchart}
\end{figure*}

Lastly, we process the features obtained from the previous network using a ResNet, which can be considered a post-processing step and optionally interpolates/coarsens from $n_s$ to $n_x$. Denote this network as $\hat{\TM} = G(F(\TD, E (\bh,\theta_E), \theta_F), \theta_G) : \mathbb{R}^{n_{c2} \times n_z \times n_s} \rightarrow \mathbb{R}^{n_z \times n_x}$, where $\hat{\TM}$ constitutes the final conductivity model estimate. The full architecture is presented in Table \ref{network_design_G}, and Fig. \ref{fig:flowchart} provides a schematic overview of the whole data pipeline, from 1D inversions to data creation to network training.

\clearpage
Training of all networks jointly proceeds via minimizing the mean squared error (MSE) over $n_{ex}$ examples using the Adam algorithm \citep{Kingma2014}:
 
\begin{equation}\label{eq:loss}
    \min_{\theta_E,\theta_F,\theta_G} L(\TD,\TM,\bh,\theta_E,\theta_F,\theta_G) =  
    \frac{1}{n_{ex}} \sum_{i=1}^{n_{ex}}\| G(F(\TD_i, E (\bh_i,\theta_E), \theta_F), \theta_G) - \TM_i \|_2^2
\end{equation}
 The loss $L(\TD,\TM,\bh,\theta_E,\theta_F,\theta_G)$ simply measures the difference in $\ell_2$ sense between the labels (conductivity models) $\TM$ and the network prediction based on flight heights $\bh$ and observed data $\TD$.

 \begin{table}
\caption{Network design for merging embedded flight heights and AEM data, as well as time-to-depth conversion, $F(\TD,E (\bh,\theta_E), \theta_F)$}
\label{network_design_F}
\begin{center}
\begin{tabular}{rrrl}
\hline 
\multicolumn{1}{c}{\bf Layer} &\multicolumn{1}{c}{\bf Channels} &\multicolumn{1}{c}{\bf Feature size} &\multicolumn{1}{c}{\bf type}\\
\hline 
$\TD_1$ & 64    & $16 \times 128$             & \textbf{embedded flight heights}\\
$\TD_2$ & 2     & $16 \times 128$   & \textbf{AEM data (time-space)}\\
1       & 64    & $16 \times 128$   & $\TX_3 = N(\TK_2 \TD_2)$\\
2       & 64    & $16 \times 128$             & $\TX_4 = \TX_3 + \TD_1$ \\
3-5       & 64    & $16 \times 128$             & $\TX_{i+1} = P_t ( \TX_i + \sigma(N(\TK_i \TX_i)))$\\
6-7       & 128    & $8 \times 128$             & $\TX_{i+1} = P_t ( \TX_i + \sigma(N(\TK_i \TX_i)))$\\
8-9       & 256    & $4 \times 128$             & $\TX_{i+1} = \TX_i + \sigma(N(\TK_i \TX_i))$\\
10       & 256    & $1 \times 128$             & $\TX_{i+1} = \operatorname{mean}_t (\TX_i)$ \\
11       & 1    & $256 \times 128$             & reassign channels $\rightarrow$ depth \\

12       & 1    & $49 \times 128$             & $\TX_{i+1} = \TA \TX_i$ \\

output       & 1    & $49 \times 128$             & conductivity (depth$\times$space) \\
\hline
\end{tabular}
\end{center}
\end{table}

\begin{table}
\caption{Network design for post-processing the depth-conductivity features into the final model. This network is a ResNet with a final channel reducing convolution at the end. Up/down sampling may be included in the network to change the spatial resolution.}
\label{network_design_G}
\begin{center}
\begin{tabular}{rrrl}
\hline
\multicolumn{1}{c}{\bf Layer} &\multicolumn{1}{c}{\bf Channels} &\multicolumn{1}{c}{\bf Feature size} &\multicolumn{1}{c}{\bf Type}\\
\hline 
input & 1    & $49 \times 128$             & \\
1       & 64    & $49 \times 128$   & $\TX_{i+1} = \TK \TX_i$\\
2-6       & 64    & $49 \times 128$   & $\TX_{i+1} = \TX{i} - t \TK_i^\top \sigma(N(\TK_i \TX_i))$\\
7       & 1    & $49 \times 128$   & $\TX_{i+1} = \TK_i \TX_i$\\
\hline
\end{tabular}
\end{center}
\end{table}

\section{Training set construction}

Because our network design, training, and inference do not include explicit physics or PDE solutions, all such information must come from the training data and labels. 

Our contribution to training set generation is a novel approach that combines geostatistical modeling and 1D inversions. The aim is to generate labels as close as possible to the target real dataset.

\begin{figure}[H]
    \centering
    \includegraphics[width=0.7\linewidth]{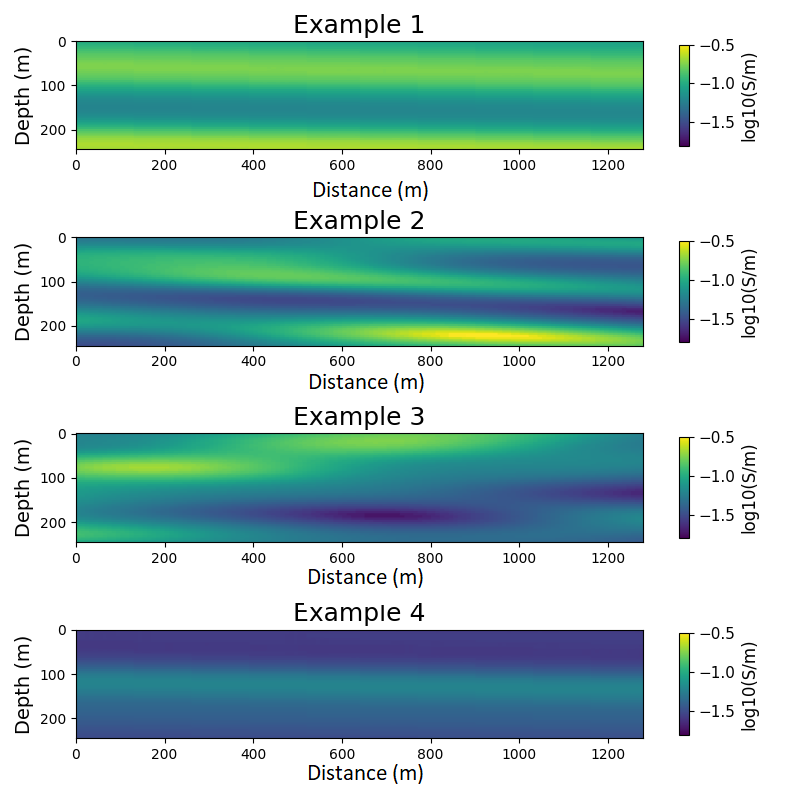}
    \caption{Four out of the $10,000$ training examples for aquifer bearing geology models.}
    \label{fig:TrainExamples}
\end{figure}

The absence of ground-truth 2D or 3D earth models is one of the primary bottlenecks of successful machine learning in the earth sciences. However, the typical subsurface structures may be known a priori for a specific area. Here, we focus on aquifer-bearing geology. Using geostatistical modeling algorithms, GSTools \citep{Muller2022}, we generate $10,000$ training models with geological structures that range from layered to smoothly varying geology and discrete anomaly models. See Fig. \ref{fig:TrainExamples} for some examples. 

\begin{figure}
	\includegraphics[width=0.8\columnwidth]{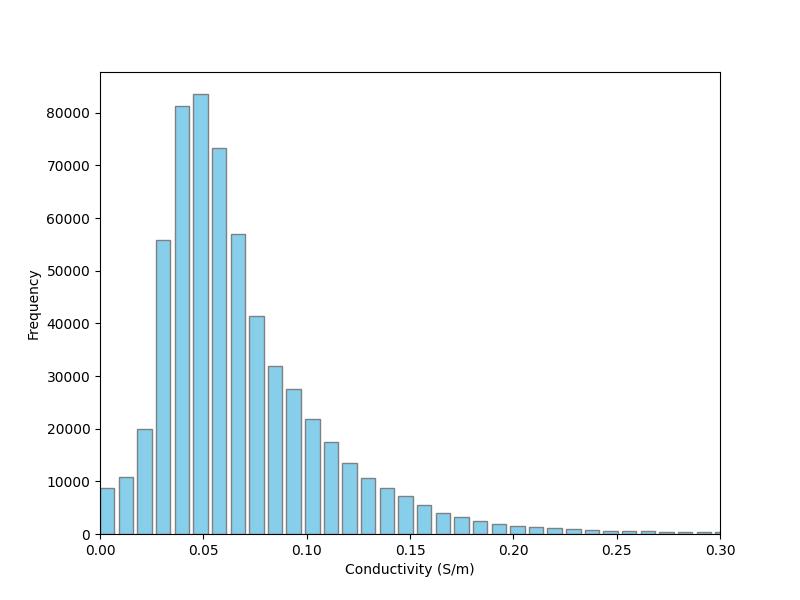}
	\caption{Histogram of conductivities derived from 1D AEM data inversion.}
	\label{fig:histogram}
\end{figure}

The training models must still be populated with realistic conductivity values to construct physical models for data generation. Here, we propose to obtain the conductivity values for our target area by employing the strengths of 1D inversion while avoiding the method's weaknesses. That is, we invert in 1D a publicly available SkyTEM dataset from the Kaweah Subbasin and use the resulting conductivity distribution to build our training models. While the 1D models are fast and cheap to compute, they will be inaccurate when encountering geological variations in 2D or 3D. Therefore, we discard the models themselves and only use the global statistics of all 1D models jointly. Fig. \ref{fig:histogram} displays the distribution of conductivities of all 1D inversion results. We assume that the conductivity values from the 1D inversions represent the global distribution of conductivities within this aquifer-bearing environment. Next, histogram matching is applied to all geostatistical models jointly to ensure that they match the global conductivity distribution of the 1D inversions. Step $1$ in Fig. \ref{fig:flowchart} summarizes the workflow detailed in this section so far.

The field area represents a large-scale underground aquifer system located in the Central Valley region of California, USA and is subject to extensive underground hydrogeological and geophysical studies \citep{knight2018mapping, Kang2022, west2007water}.

We use 3D forward modeling software (H3DTD, \cite{Haber2007}) to compute the electromagnetic fields and simulate observed data from our 2D models, extended to 3D as in a 2.5D modeling approach. Transmitter waveforms for high and low-moments and other data acquisition specifications are matched to the SkyTEM system. Lastly, our neural network needs realistic flight height patterns to generalize to field data. Therefore, we create the synthetic training flight heights by extracting the transmitter heights from the flight lines of the real SkyTEM dataset and then we assign random subsets of the lines to the 2D training models. The synthetic data generation corresponds to step $2$ in Fig. \ref{fig:flowchart}.

The conductivity models and flight heights use the dataset's global standard deviation and mean for basic normalization prior to training. In contrast, the AEM data uses only the standard deviation from each flight line for normalization, meaning the data in each flight line is divided by the standard deviation within the given flight line for each time gate. By not dividing by the mean, the global scale of the data is preserved, which is important for differentiating between various conductive and resistive anomalies.

\section{Results}

\begin{figure}
	\centering
	\includegraphics[width=0.65\columnwidth]{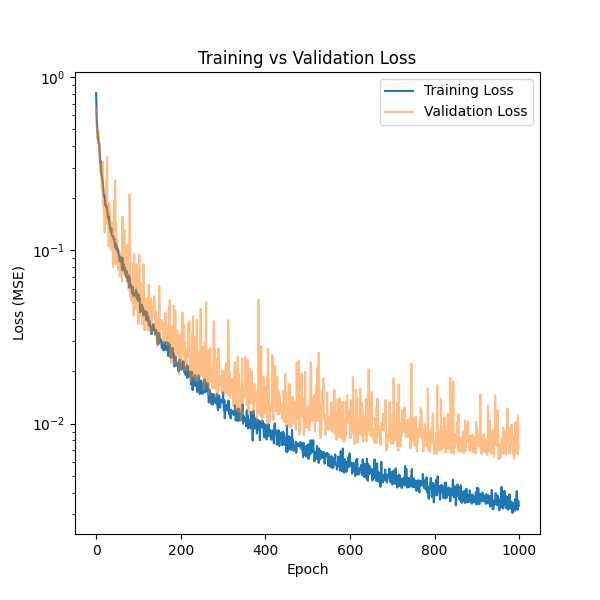}
	\caption{Training and validation loss.}
	\label{fig:epochLoss}
\end{figure}

We present results from artificial intelligence (AI) predictions using the network trained via eq.~\eqref{eq:loss}. Training included basic data augmentation on the $10,000$ samples by flipping each sample horizontally, and training and validation sets used a random $90\%/10 \%$ split of the full data set. Training is fast: even using a relatively old NVIDIA GPU Geforce RTX 2070, the network takes $\approx$ a day to train. See Fig. \ref{fig:epochLoss} for training and validation loss histories. We stopped training once the validation loss reached an approximately stationary value. For AEM data, it is entirely expected that the validation loss stalls while the training loss keeps decreasing. The reason is the non-uniqueness of the data. Many conductivity models generate almost the same data. A prediction from that data cannot uniquely determine which model is the `true' model. Once the network is trained, inference provides conductivity models almost instantaneously.

\begin{figure}
	\centering
	\includegraphics[width=0.8\linewidth]{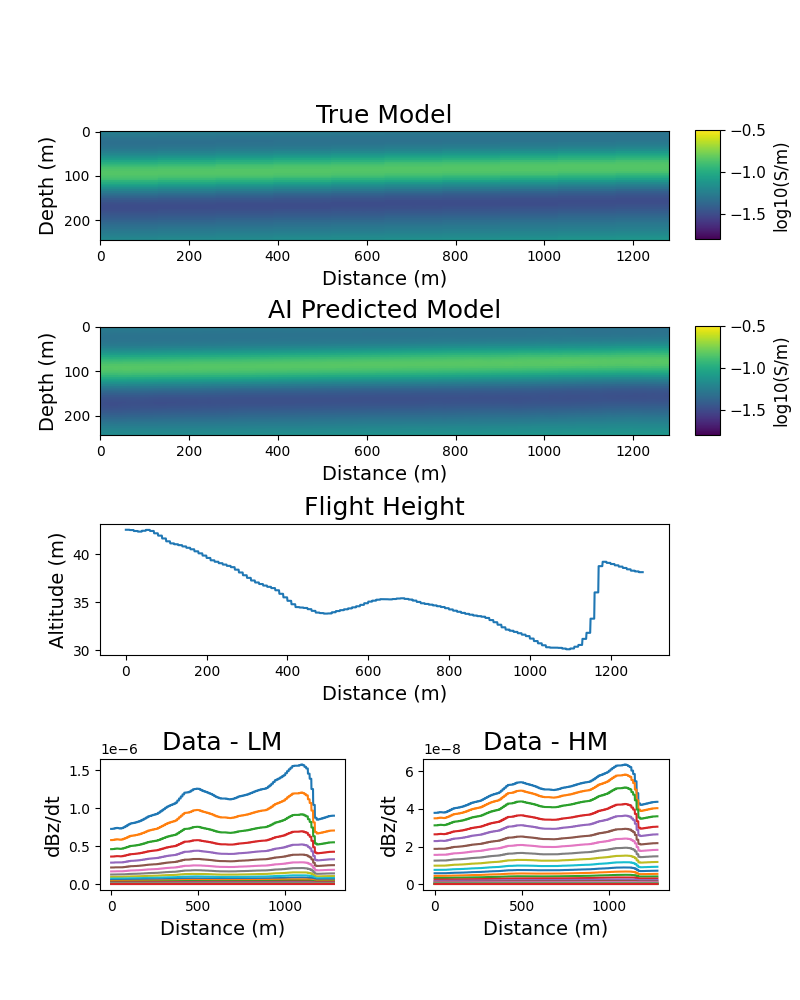}
	\caption{An example from the validation set with the true model, AI predicted model, flight heights, and synthetic high and low-moment data.}
	\label{fig:valComparisons}
\end{figure}

The low validation loss suggests accurate and visually near-perfect model reconstruction errors on the training and validation sets, as seen in Fig. \ref{fig:valComparisons} along with the flight height and forward modeled synthetic high and low moment data. This figure illustrates that the network can predict the conductivity model with near-perfect accuracy, as is desired with validation data.

Good performance on the validation set is necessary but not sufficient for performance on the field dataset, which is considered our test set. We apply our trained network as-is, without further modifications, transfer learning, or fine-tuning. All work for synthetic conductivity-model generation, 1D inversion, 3D forward modeling, and training was done beforehand. 

\begin{figure}
   \centering
   \begin{subfigure}{0.9\linewidth}
   \includegraphics[width=0.8\linewidth]{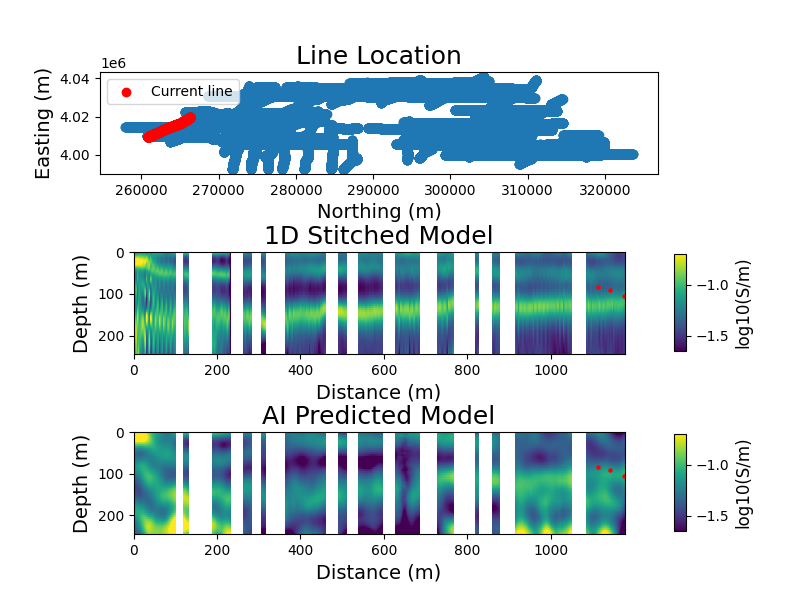}
   \caption{Line 100102}
  \end{subfigure}
   \begin{subfigure}{0.9\linewidth}
    \includegraphics[width=0.8\linewidth]{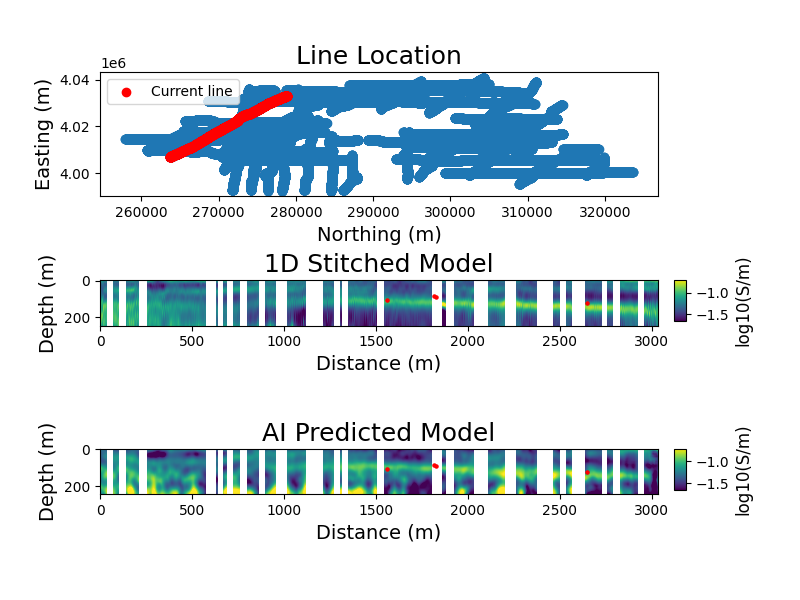}
    \caption{Line 100203}
  \end{subfigure}
   \caption{Two flight lines from Kaweah (California) data, our neural network prediction (log conductivity) and the 1D stitched inversion for comparison. Masked areas do not contain observed data within 30m of the location. Red dots are tops of Corcoran clay depths as found in surrounding drill holes within 5km of the line.}
   \label{fig:MaskedPred}
\end{figure}

Fig.~\ref{fig:MaskedPred} shows the 1D stitched models from two field lines at Kaweah along with our AI prediction models, where the models are masked where observed data is missing. Note that when data is missing, the gaps are filled in with interpolation before predicting a conductivity model for the entire area. The AI models are only masked after the whole prediction has been completed. The stitched 1D models look plausible, albeit with some 1D artifacts, which could be mitigated using lateral regularization. The AI predictions have similar-looking features to the 1D results, which suggests that the network is predicting geologically plausible models. This implies that the observed data is sufficiently `close' to the synthetic training data for the network to `recognize' it. To further evaluate the results, the red dots on Fig.~\ref{fig:MaskedPred} represent locations of the top of the conductive Corcoran clay layer as found in nearby boreholes. This thin 10-20 meter thick clay layer separates an upper and a lower aquifer region \citep{Kang2022} and is an important geologic feature to understand in order to map the aquifer structures in the area. Fig.~\ref{fig:MaskedPred}a shows that the Corcoran clay layer maps well to the AI inversion model but not as well to the 1D stitched model. In Fig.~\ref{fig:MaskedPred}b there is approximately an equally good correlation between the Corcoran clay drillhole markers and both the 1D stitched model and the AI predicted model. This successful comparison with drillhole information is an important step in helping to validate the AI prediction process.

\begin{figure*}
    \centering
    {\includegraphics[width=0.45\linewidth]{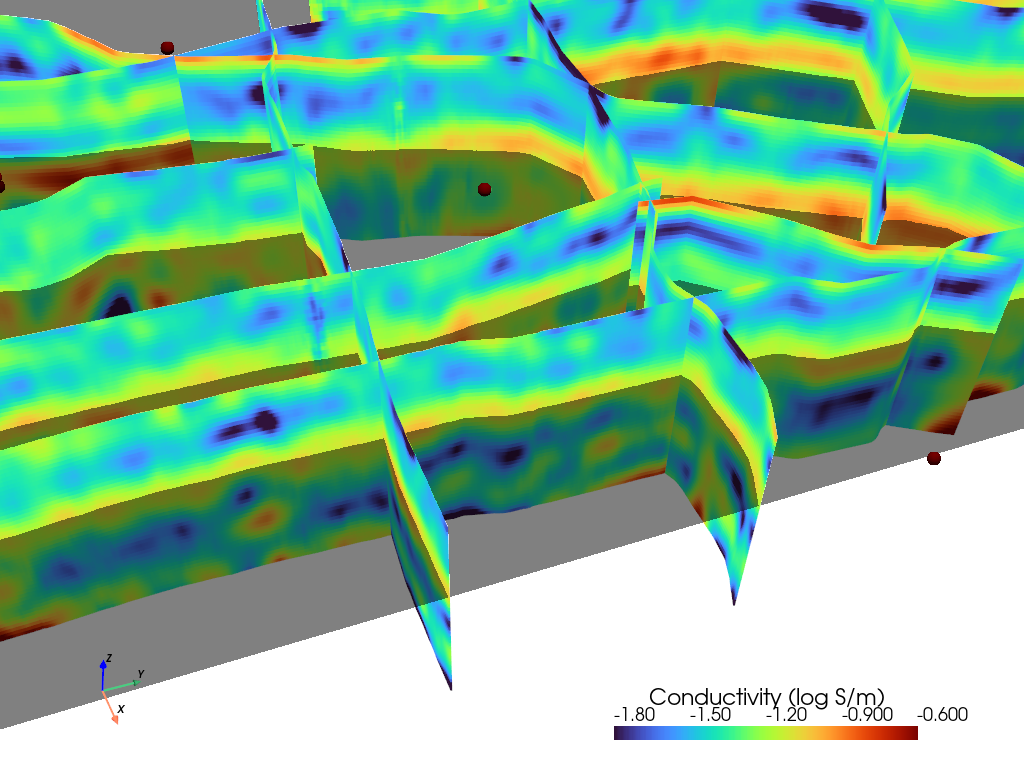}}
    {\includegraphics[width=0.45\linewidth]{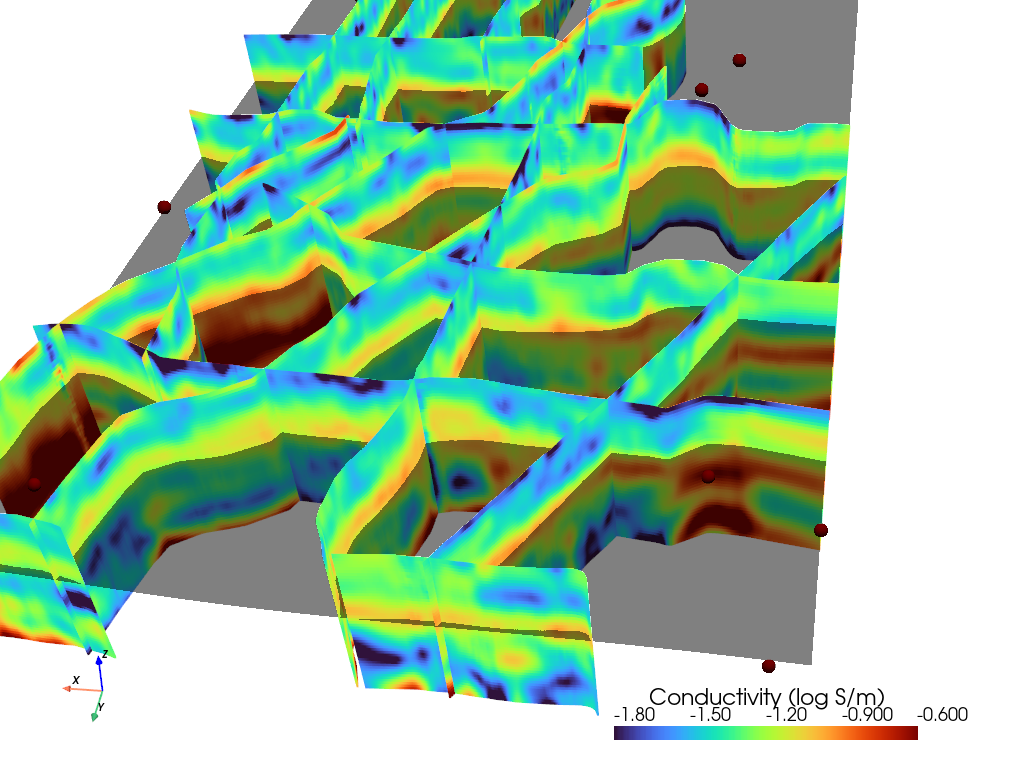}} 
    \caption{3D fence plots for the AI predicted conductivity models for the Kaweah (California) dataset together with the interpolated Corcoran clay layer from drill holes. True drill hole locations are shown as red dots. a) Looking West. b) Looking South.}
    \label{fig:Pred3D}
\end{figure*}

Next, we look at the predictions for all flight lines together. This results in the 3D fence diagrams in Fig. \ref{fig:Pred3D}. The AI predictions are independent for each flight line and should result in spatially coherent conductivity models that match at the intersections of flight lines. Fig. \ref{fig:Pred3D} shows continuity at the intersections, which gives us confidence that the network predicts conductivity models that are consistent with each other. This means that the models satisfy the prior knowledge that was implicitly included via the synthetically generated training models. Fig.~\ref{fig:Pred3D} also shows the top of the Corcoran clay layer as a transparent grey surface, as interpolated between drillhole values. Fig.~\ref{fig:Pred3D} shows that the Corcoran clay layer has a good correlation with the conductive layer in the AI prediction model for most lines, although there are some lines where the Corcoran clay appears slightly below the AI predicted conductor. It is known that this clay layer is only 10-20m thick, so the AI predictions are also perhaps too thick in some spots, thus emphasizing the need for thinner conductive layers in the training models. But overall, there is a good relationship between the Corcoran clay observed in drilling and the conductive layer in the AI predictions.

\begin{figure}
    \centering
    \includegraphics[height=0.5\textheight, keepaspectratio, width=0.8\linewidth]{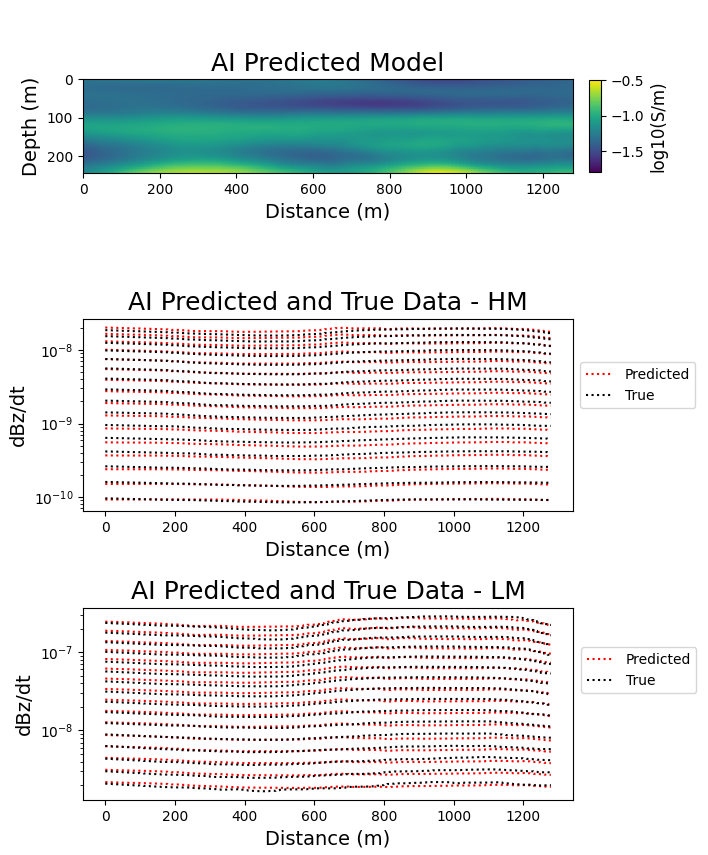}
    \caption{AI predicted model for a flight line at Kaweah, along with observed and predicted data plots for both high-moment (HM) and low-moment (LM) data. Predicted data consists of 3D forward modeled data from extending the 2D AI prediction model into a 3D model.} 
    \label{fig:FWD_on_Pred}
\end{figure}

Lastly, we display how well the AI predicted models match the governing physics by showing the observed data and forward modeled predicted data. We calculate the predicted data corresponding to our AI models at the cost of one 3D forward modeling per station per waveform. The cell size in the core region of the forward mesh is $10$m $\times$ $10$m $\times$ $10$m in $x$, $y$ and $z$, respectively. Fig. \ref{fig:FWD_on_Pred} shows this comparison for one flight line at Kaweah, where the AI predicted model is shown at the top followed by the observed and predicted data for the high-moment and low-moment data below. This figure shows a close match between the observed data and the forward-modeled data from our AI predicted models. This demonstrates that our approach not only generates geologically feasible models that match drillhole observations but that these models also obey the governing physics.

\section{Discussion}
The results show that our AI predictions satisfy both the prior knowledge and the observed data, which leads to the conclusion that the network provides reasonable earth conductivity models. Our case study in the Kaweah Subbasin verifies the applicability of the presented methodology to groundwater investigations within an aquifer-bearing region. We note that the inverse problem's intrinsic non-uniqueness remains, and other conductivity models may satisfy the prior knowledge and observed data. However, we severely limit the variety of models that the network can predict by generating earth structures for a specific geological setting with conductivity distributions obtained from 1D inversions. This is, in effect, a strong regularizer, or narrow prior, applied to the inverse problem.

Our approach is bootstrapped because 1D inversions are used to estimate the distribution of conductivity values for training model generation. The `structure' comes from prior knowledge of the local geology. Because these 1D inversions and 3D forward modeling are a one-time expense ahead of training, the presented network also applies to real-time inversion, provided that new data comes from the training area or a similar region. Applying the method to different geological settings, e.g., mineral exploration, requires its own set of geostatistical models, new 1D inversions and subsequent 3D forward modeling of synthetic models. Looking forward, another level of bootstrapping would entail generating more specific training models based on the predictions so far and drillhole information, e.g., including thin clay layers. It should be noted that validating such an approach will become more challenging because the drillhole information is then used for training and thus has lost its validation powers. 

\section{Conclusions}

This study introduces a novel approach for inverting airborne electromagnetic (AEM) data using a machine-learning framework tailored to geological scenarios with specific characteristics. By leveraging a combination of fast 1D inversions with synthetic geostatistical modeling to generate training data for a custom neural network, the proposed method effectively addresses the computational challenges of traditional 3D inversion. The method also produces geologically plausible conductivity models without the need for forward modeling during inference. The approach has been validated on the Kaweah Subbasin SkyTEM AEM dataset, demonstrating its ability to generate spatially coherent models in an aquifer-bearing region that aligns with both observed data and prior geological knowledge.

For the first time, we demonstrated the feasibility of predicting full 2D conductivity models from AEM data along with flight height information while generalizing to a real dataset. This highlights the potential of this approach for large-scale groundwater investigations and other subsurface exploration, possibly in real time.

\clearpage

\normalsize
\bibliography{paper_bibliography}

\begin{thebibliography}{54}
\providecommand{\natexlab}[1]{#1}
\providecommand{\url}[1]{\texttt{#1}}
\expandafter\ifx\csname urlstyle\endcsname\relax
  \providecommand{\doi}[1]{doi: #1}\else
  \providecommand{\doi}{doi: \begingroup \urlstyle{rm}\Url}\fi

\bibitem[Viezzoli et~al.(2010)Viezzoli, Munday, Auken, Christiansen, and Wilson]{viezzoli2010accurate}
Andrea Viezzoli, Tim Munday, Esben Auken, Anders~V Christiansen, and Glenn~A Wilson.
\newblock Accurate quasi 3d versus practical full 3d inversion of aem data--the bookpurnong case study.
\newblock \emph{Preview}, 2010\penalty0 (149):\penalty0 23--31, 2010.

\bibitem[Leif H.~Cox and Zhdanov(2010)]{doi:10.1071/EG10003}
Glenn A.~Wilson Leif H.~Cox and Michael~S. Zhdanov.
\newblock 3d inversion of airborne electromagnetic data using a moving footprint.
\newblock \emph{Exploration Geophysics}, 41\penalty0 (4):\penalty0 250--259, 2010.
\newblock \doi{10.1071/EG10003}.
\newblock URL \url{https://doi.org/10.1071/EG10003}.

\bibitem[Minsley(2011)]{MinsleyTDMC}
Burke~J. Minsley.
\newblock {A trans-dimensional Bayesian Markov chain Monte Carlo algorithm for model assessment using frequency-domain electromagnetic data}.
\newblock \emph{Geophysical Journal International}, 187\penalty0 (1):\penalty0 252--272, 10 2011.
\newblock ISSN 0956-540X.
\newblock \doi{10.1111/j.1365-246X.2011.05165.x}.
\newblock URL \url{https://doi.org/10.1111/j.1365-246X.2011.05165.x}.

\bibitem[Abraham et~al.(2012)Abraham, Minsley, Bedrosian, Smith, and Cannia]{AbrahamAEMwater}
Jared~D. Abraham, Burke~J. Minsley, Paul~A. Bedrosian, Bruce~D. Smith, and James~C. Cannia.
\newblock Airborne electromagnetic surveys for groundwater characterization.
\newblock \emph{ASEG Extended Abstracts}, 2012\penalty0 (1):\penalty0 1--4, 2012.
\newblock \doi{10.1071/ASEG2012ab246}.
\newblock URL \url{https://doi.org/10.1071/ASEG2012ab246}.

\bibitem[Bedrosian et~al.(2015)Bedrosian, Schamper, and Auken]{BedrosianComparison}
Paul~A. Bedrosian, Cyril Schamper, and Esben Auken.
\newblock A comparison of helicopter‐borne electromagnetic systems for hydrogeologic studies.
\newblock \emph{Geophysical Prospecting}, 64\penalty0 (1):\penalty0 192--215, 2015.
\newblock ISSN 1365-2478.
\newblock \doi{https://doi.org/10.1111/1365-2478.12262}.
\newblock URL \url{https://www.earthdoc.org/content/journals/10.1111/1365-2478.12262}.

\bibitem[McMillan et~al.(2018)McMillan, Haber, and Marchant]{McMillan2018}
Michael~Stanley McMillan, Eldad Haber, and Dave Marchant.
\newblock Large scale 3d airborne electromagnetic inversion–recent technical improvements.
\newblock \emph{ASEG Extended Abstracts}, pages 1--6, 2018.

\bibitem[McMillan et~al.(2019)McMillan, Haber, and Lawrie]{McMillan2019}
Michael~Stanley McMillan, Eldad Haber, and Ken Lawrie.
\newblock Noise, constraints and 3d inversion: A practical look at high-resolution aquifer mapping using airborne electromagnetics.
\newblock \emph{SEG International Exposition and Annual Meeting}, 2019.

\bibitem[Kang et~al.(2022)Kang, Knight, and Goebel]{Kang2022}
Seogi Kang, Rosemary Knight, and Meredith Goebel.
\newblock Improved imaging of the large-scale structure of a groundwater system with airborne electromagnetic data.
\newblock \emph{Water Resources Research}, 58, 4 2022.
\newblock ISSN 19447973.
\newblock \doi{10.1029/2021WR031439}.

\bibitem[Knight et~al.(2022)Knight, Steklova, Miltenberger, Kang, Goebel, and Fogg]{Knight_2022}
Rosemary Knight, Klara Steklova, Alex Miltenberger, Seogi Kang, Meredith Goebel, and Graham Fogg.
\newblock Airborne geophysical method images fast paths for managed recharge of california’s groundwater.
\newblock \emph{Environmental Research Letters}, 17\penalty0 (12):\penalty0 124021, dec 2022.
\newblock \doi{10.1088/1748-9326/aca344}.
\newblock URL \url{https://dx.doi.org/10.1088/1748-9326/aca344}.

\bibitem[Christensen et~al.(2017)Christensen, Minsley, and Christensen]{Christensen3DHydro}
N.~K. Christensen, B.~J. Minsley, and S.~Christensen.
\newblock Generation of 3-d hydrostratigraphic zones from dense airborne electromagnetic data to assess groundwater model prediction error.
\newblock \emph{Water Resources Research}, 53\penalty0 (2):\penalty0 1019--1038, 2017.
\newblock \doi{https://doi.org/10.1002/2016WR019141}.
\newblock URL \url{https://agupubs.onlinelibrary.wiley.com/doi/abs/10.1002/2016WR019141}.

\bibitem[Zhang(2003)]{Zhang3DEM}
Zhiyi Zhang.
\newblock 3d resistivity mapping of airborne em data.
\newblock \emph{GEOPHYSICS}, 68\penalty0 (6):\penalty0 1896--1905, 2003.
\newblock \doi{10.1190/1.1635042}.
\newblock URL \url{https://doi.org/10.1190/1.1635042}.

\bibitem[Haber et~al.(2007)Haber, Oldenburg, and Shekhtman]{Haber2007}
Eldad Haber, Douglas~W. Oldenburg, and R.~Shekhtman.
\newblock Inversion of time domain three-dimensional electromagnetic data.
\newblock \emph{Geophysical Journal International}, 171:\penalty0 550--564, 11 2007.
\newblock ISSN 0956540X.
\newblock \doi{10.1111/j.1365-246X.2007.03365.x}.
\newblock URL \url{http://doi.wiley.com/10.1111/j.1365-246X.2007.03365.x}.

\bibitem[Cox et~al.(2012)Cox, Wilson, and Zhdanov]{doi:10.1190/geo2011-0370.1}
Leif~H. Cox, Glenn~A. Wilson, and Michael~S. Zhdanov.
\newblock 3d inversion of airborne electromagnetic data.
\newblock \emph{GEOPHYSICS}, 77\penalty0 (4):\penalty0 WB59--WB69, 2012.
\newblock \doi{10.1190/geo2011-0370.1}.
\newblock URL \url{https://doi.org/10.1190/geo2011-0370.1}.

\bibitem[McMillan et~al.(2015)McMillan, Schwarzbach, Haber, and Oldenburg]{McMillan2015}
Michael~S. McMillan, Christoph Schwarzbach, Eldad Haber, and Douglas~W. Oldenburg.
\newblock 3d parametric hybrid inversion of time-domain airborne electromagnetic data.
\newblock \emph{Geophysics}, 80:\penalty0 K25--K36, 9 2015.
\newblock ISSN 0016-8033.
\newblock \doi{10.1190/geo2015-0141.1}.
\newblock URL \url{http://library.seg.org/doi/10.1190/geo2015-0141.1}.

\bibitem[Ren et~al.(2020)Ren, Macnae, and Hennessy]{ren2020three}
Xiuyan Ren, James Macnae, and Lachlan Hennessy.
\newblock Three conductivity modelling algorithms and three 3d inversions of the forrestania test site aem anomaly.
\newblock \emph{Exploration Geophysics}, 51\penalty0 (1):\penalty0 14--24, 2020.

\bibitem[Farquharson and Oldenburg(1996)]{Farquharson1996}
C.~G. Farquharson and D.~W. Oldenburg.
\newblock Approximate sensitivities for the electromagnetic inverse problem.
\newblock \emph{Geophysical Journal International}, 126:\penalty0 235--252, 7 1996.
\newblock ISSN 0956540X.
\newblock \doi{10.1111/j.1365-246X.1996.tb05282.x}.
\newblock URL \url{http://gji.oxfordjournals.org/cgi/doi/10.1111/j.1365-246X.1996.tb05282.x}.

\bibitem[Tartaras and Beamish(2006)]{Lat1DInvFixedWing}
E.~Tartaras and D.~Beamish.
\newblock Laterally constrained inversion of fixed-wing frequency-domain aem data.
\newblock art. cp-14-00003, 2006.
\newblock ISSN 2214-4609.
\newblock \doi{https://doi.org/10.3997/2214-4609.201402625}.
\newblock URL \url{https://www.earthdoc.org/content/papers/10.3997/2214-4609.201402625}.

\bibitem[Viezzoli et~al.(2008)Viezzoli, Christiansen, Auken, and Sørensen]{viezzoli2008spatially}
Andrea Viezzoli, Anders~Vest Christiansen, Esben Auken, and Kurt Sørensen.
\newblock Quasi-3d modeling of airborne tem data by spatially constrained inversion.
\newblock \emph{GEOPHYSICS}, 73\penalty0 (3):\penalty0 F105--F113, 2008.
\newblock \doi{10.1190/1.2895521}.
\newblock URL \url{https://doi.org/10.1190/1.2895521}.

\bibitem[Brodie and Sambridge(2012)]{brodie2012transdimensional}
Ross~C Brodie and Malcolm Sambridge.
\newblock Transdimensional monte carlo inversion of aem data.
\newblock \emph{ASEG Extended Abstracts}, 2012\penalty0 (1):\penalty0 1--4, 2012.

\bibitem[Wolfgram et~al.(2003)Wolfgram, Sattel, and Christensen]{wolfgram2003approximate}
Peter Wolfgram, Daniel Sattel, and Niels~B Christensen.
\newblock Approximate 2d inversion of aem data.
\newblock \emph{Exploration Geophysics}, 34\penalty0 (2):\penalty0 29--33, 2003.

\bibitem[Wilson et~al.(2006)Wilson, Raiche, and Sugeng]{Wilson2006}
G.~A. Wilson, A.~P. Raiche, and F.~Sugeng.
\newblock 2.5d inversion of airborne electromagnetic data.
\newblock \emph{Exploration Geophysics}, 37:\penalty0 363--371, 2006.
\newblock ISSN 18347533.
\newblock \doi{10.1071/EG06363}.

\bibitem[J.~Silic and FitzGerald(2018)]{inv1Dvs25D}
R.~Paterson J.~Silic and D.~FitzGerald.
\newblock 2.5d vs 1d aem forward and inversion methods at a survey scale : A case study.
\newblock \emph{ASEG Extended Abstracts}, 2018\penalty0 (1):\penalty0 1--8, 2018.
\newblock \doi{10.1071/ASEG2018abM2\_1E}.
\newblock URL \url{https://doi.org/10.1071/ASEG2018abM2_1E}.

\bibitem[Ellis(1998)]{ellis1998inversion}
Robert~G Ellis.
\newblock Inversion of airborne electromagnetic data.
\newblock \emph{Exploration Geophysics}, 29\penalty0 (1-2):\penalty0 121--127, 1998.

\bibitem[Lin et~al.(2019)Lin, Fiandaca, Auken, Couto, and Christiansen]{lin2019discussion}
Changhong Lin, Gianluca Fiandaca, Esben Auken, Marco~Antonio Couto, and Anders~Vest Christiansen.
\newblock A discussion of 2d induced polarization effects in airborne electromagnetic and inversion with a robust 1d laterally constrained inversion scheme.
\newblock \emph{Geophysics}, 84\penalty0 (2):\penalty0 E75--E88, 2019.

\bibitem[Deleersnyder et~al.(2022)Deleersnyder, Dudal, and Hermans]{deleersnyder2022assessing}
Wouter Deleersnyder, David Dudal, and Thomas Hermans.
\newblock Assessing quantitatively interpretable zones from 1d forward modelling aem inversion models.
\newblock In \emph{NSG2022 3rd Conference on Airborne, Drone and Robotic Geophysics}, volume 2022, pages 1--5. European Association of Geoscientists \& Engineers, 2022.

\bibitem[Hamilton and Hauptmann(2018)]{hamilton2018deep}
Sarah~Jane Hamilton and Andreas Hauptmann.
\newblock Deep d-bar: Real-time electrical impedance tomography imaging with deep neural networks.
\newblock \emph{IEEE transactions on medical imaging}, 37\penalty0 (10):\penalty0 2367--2377, 2018.

\bibitem[Sainsbury-Dale et~al.(2022)Sainsbury-Dale, Zammit-Mangion, and Huser]{sainsbury2022fast}
Matthew Sainsbury-Dale, Andrew Zammit-Mangion, and Rapha{\"e}l Huser.
\newblock Fast optimal estimation with intractable models using permutation-invariant neural networks.
\newblock \emph{arXiv preprint arXiv:2208.12942}, 2022.

\bibitem[Deng et~al.(2022)Deng, Feng, Wang, Zhang, Jin, Feng, Zeng, Chen, and Lin]{deng2022openfwi}
Chengyuan Deng, Shihang Feng, Hanchen Wang, Xitong Zhang, Peng Jin, Yinan Feng, Qili Zeng, Yinpeng Chen, and Youzuo Lin.
\newblock Openfwi: Large-scale multi-structural benchmark datasets for full waveform inversion.
\newblock \emph{Advances in Neural Information Processing Systems}, 35:\penalty0 6007--6020, 2022.

\bibitem[Chung et~al.(2024)Chung, Hart, Chung, Peters, and Haber]{chung2024paired}
Matthias Chung, Emma Hart, Julianne Chung, Bas Peters, and Eldad Haber.
\newblock Paired autoencoders for likelihood-free estimation in inverse problems.
\newblock \emph{Machine Learning: Science and Technology}, 5\penalty0 (4):\penalty0 045055, dec 2024.
\newblock \doi{10.1088/2632-2153/ad95dd}.
\newblock URL \url{https://dx.doi.org/10.1088/2632-2153/ad95dd}.

\bibitem[Asif et~al.(2022)Asif, Foged, Maurya, Grombacher, Christiansen, Auken, and Larsen]{asif2022integrating}
Muhammad~Rizwan Asif, Nikolaj Foged, Pradip~Kumar Maurya, Denys~James Grombacher, Anders~Vest Christiansen, Esben Auken, and Jakob~Juul Larsen.
\newblock Integrating neural networks in least-squares inversion of airborne time-domain electromagnetic data.
\newblock \emph{Geophysics}, 87\penalty0 (4):\penalty0 E177--E187, 2022.

\bibitem[Wu et~al.(2022)Wu, Huang, and Zhao]{Wu2022}
Sihong Wu, Qinghua Huang, and Li~Zhao.
\newblock Instantaneous inversion of airborne electromagnetic data based on deep learning.
\newblock \emph{Geophysical Research Letters}, 49, 5 2022.
\newblock ISSN 19448007.
\newblock \doi{10.1029/2021GL097165}.

\bibitem[Bai et~al.(2020)Bai, Vignoli, Viezzoli, Nevalainen, and Vacca]{rs12203440}
Peng Bai, Giulio Vignoli, Andrea Viezzoli, Jouni Nevalainen, and Giuseppina Vacca.
\newblock (quasi-)real-time inversion of airborne time-domain electromagnetic data via artificial neural network.
\newblock \emph{Remote Sensing}, 12\penalty0 (20), 2020.
\newblock ISSN 2072-4292.
\newblock \doi{10.3390/rs12203440}.
\newblock URL \url{https://www.mdpi.com/2072-4292/12/20/3440}.

\bibitem[Feng et~al.(2020)Feng, feng Zhang, Li, and Bai]{Bing1DNN}
Bing Feng, Ji~feng Zhang, Dong Li, and Yang Bai.
\newblock Resistivity-depth imaging with the airborne transient electromagnetic method based on an artificial neural network.
\newblock \emph{Journal of Environmental and Engineering Geophysics}, 25\penalty0 (3):\penalty0 355--368, 2020.
\newblock \doi{10.32389/JEEG19-087}.
\newblock URL \url{https://doi.org/10.32389/JEEG19-087}.

\bibitem[Li et~al.(2020{\natexlab{a}})Li, Liu, Yin, Ren, and Su]{Li1DDNN}
Jinfeng Li, Yunhe Liu, Changchun Yin, Xiuyan Ren, and Yang Su.
\newblock Fast imaging of time-domain airborne em data using deep learning technology.
\newblock \emph{GEOPHYSICS}, 85\penalty0 (5):\penalty0 E163--E170, 2020{\natexlab{a}}.
\newblock \doi{10.1190/geo2019-0015.1}.
\newblock URL \url{https://doi.org/10.1190/geo2019-0015.1}.

\bibitem[Kyubo~Noh and Byun(2020)]{Noh1DDNN}
Daeung~Yoon Kyubo~Noh and Joongmoo Byun.
\newblock Imaging subsurface resistivity structure from airborne electromagnetic induction data using deep neural network.
\newblock \emph{Exploration Geophysics}, 51\penalty0 (2):\penalty0 214--220, 2020.
\newblock \doi{10.1080/08123985.2019.1668240}.
\newblock URL \url{https://doi.org/10.1080/08123985.2019.1668240}.

\bibitem[Bang et~al.(2022)Bang, Byun, and Seol]{bang2022deep}
Minkyu Bang, Joongmoo Byun, and Soon~Jee Seol.
\newblock Deep neural network-based airborne em data inversion suitable for mountainous field sites.
\newblock In \emph{83rd EAGE Conference and Exhibition 2022}, volume~2, pages 1288--1292. European Association of Geoscientists and Engineers, EAGE, 2022.

\bibitem[Kang et~al.(2024)Kang, Bang, Seol, and Byun]{doi:10.1190/geo2022-0723.1}
Hyeonwoo Kang, Minkyu Bang, Soon~Jee Seol, and Joongmoo Byun.
\newblock Deep-learning-based airborne transient electromagnetic inversion providing the depth of investigation.
\newblock \emph{GEOPHYSICS}, 89\penalty0 (2):\penalty0 E31--E45, 2024.
\newblock \doi{10.1190/geo2022-0723.1}.
\newblock URL \url{https://doi.org/10.1190/geo2022-0723.1}.

\bibitem[Wu et~al.(2023)Wu, Huang, and Zhao]{Wu2023fastBayesian}
Sihong Wu, Qinghua Huang, and Li~Zhao.
\newblock Fast bayesian inversion of airborne electromagnetic data based on the invertible neural network.
\newblock \emph{IEEE Transactions on Geoscience and Remote Sensing}, 61:\penalty0 1--11, 2023.
\newblock \doi{10.1109/TGRS.2023.3264777}.

\bibitem[Wu et~al.(2024)Wu, Huang, and Zhao]{10.1093/gji/ggae244}
Sihong Wu, Qinghua Huang, and Li~Zhao.
\newblock Physics-guided deep learning-based inversion for airborne electromagnetic data.
\newblock \emph{Geophysical Journal International}, 238\penalty0 (3):\penalty0 1774--1789, 07 2024.
\newblock ISSN 1365-246X.
\newblock \doi{10.1093/gji/ggae244}.
\newblock URL \url{https://doi.org/10.1093/gji/ggae244}.

\bibitem[Bang et~al.(2021)Bang, Oh, Noh, Seol, and Byun]{AEMRNN}
Minkyu Bang, Seokmin Oh, Kyubo Noh, Soon~Jee Seol, and Joongmoo Byun.
\newblock Imaging subsurface orebodies with airborne electromagnetic data using a recurrent neural network.
\newblock \emph{GEOPHYSICS}, 86\penalty0 (6):\penalty0 E407--E419, 2021.
\newblock \doi{10.1190/geo2020-0871.1}.
\newblock URL \url{https://doi.org/10.1190/geo2020-0871.1}.

\bibitem[Puzyrev and Swidinsky(2021)]{PUZYREV2021104681}
Vladimir Puzyrev and Andrei Swidinsky.
\newblock Inversion of 1d frequency- and time-domain electromagnetic data with convolutional neural networks.
\newblock \emph{Computers \& Geosciences}, 149:\penalty0 104681, 2021.
\newblock ISSN 0098-3004.
\newblock \doi{https://doi.org/10.1016/j.cageo.2020.104681}.
\newblock URL \url{https://www.sciencedirect.com/science/article/pii/S009830042030652X}.

\bibitem[Puzyrev(2019)]{puzyrev2019deep}
Vladimir Puzyrev.
\newblock Deep learning electromagnetic inversion with convolutional neural networks.
\newblock \emph{Geophysical Journal International}, 218\penalty0 (2):\penalty0 817--832, 2019.

\bibitem[Li et~al.(2022)Li, Zhang, Xing, and Zheng]{9733939}
Shiyan Li, Xiaojuan Zhang, Kang Xing, and Yaoxin Zheng.
\newblock Fast inversion of subsurface target electromagnetic induction response with deep learning.
\newblock \emph{IEEE Geoscience and Remote Sensing Letters}, 19:\penalty0 1--5, 2022.
\newblock \doi{10.1109/LGRS.2022.3159269}.

\bibitem[Zhao et~al.(2024)Zhao, Wu, Chen, Xue, and Shi]{groundTEMDNN}
Yang Zhao, Xin Wu, Weiying Chen, Junjie Xue, and Jinjing Shi.
\newblock {Three-dimensional inversion for short-offset transient electromagnetic data based on 3D U-Net}.
\newblock \emph{Journal of Geophysics and Engineering}, 21\penalty0 (3):\penalty0 922--937, 04 2024.
\newblock ISSN 1742-2132.
\newblock \doi{10.1093/jge/gxae046}.
\newblock URL \url{https://doi.org/10.1093/jge/gxae046}.

\bibitem[S{\o}rense and Auken(2004)]{sorense2004skytem}
KI~S{\o}rense and Esben Auken.
\newblock Skytem? a new high-resolution helicopter transient electromagnetic system.
\newblock \emph{Exploration Geophysics}, 35\penalty0 (3):\penalty0 194--202, 2004.

\bibitem[He et~al.(2016)He, Zhang, Ren, and Sun]{He2016}
Kaiming He, Xiangyu Zhang, Shaoqing Ren, and Jian Sun.
\newblock Deep residual learning for image recognition.
\newblock \emph{Proceedings of the IEEE conference on computer vision and pattern recognition}, pages 770--778, 2016.
\newblock URL \url{http://image-net.org/challenges/LSVRC/2015/}.

\bibitem[Ronneberger et~al.(2015)Ronneberger, Fischer, and Brox]{Ronneberger2015}
Olaf Ronneberger, Philipp Fischer, and Thomas Brox.
\newblock U-net: Convolutional networks for biomedical image segmentation.
\newblock \emph{International Conference on Medical Image Computing and Computer-Assisted Intervention}, pages 234--241, 5 2015.
\newblock URL \url{http://arxiv.org/abs/1505.04597}.

\bibitem[Li et~al.(2020{\natexlab{b}})Li, Kovachki, Azizzadenesheli, Liu, Stuart, Bhattacharya, and Anandkumar]{NEURIPS2020_4b21cf96}
Zongyi Li, Nikola Kovachki, Kamyar Azizzadenesheli, Burigede Liu, Andrew Stuart, Kaushik Bhattacharya, and Anima Anandkumar.
\newblock Multipole graph neural operator for parametric partial differential equations.
\newblock In H.~Larochelle, M.~Ranzato, R.~Hadsell, M.F. Balcan, and H.~Lin, editors, \emph{Advances in Neural Information Processing Systems}, volume~33, pages 6755--6766. Curran Associates, Inc., 2020{\natexlab{b}}.
\newblock URL \url{https://proceedings.neurips.cc/paper_files/paper/2020/file/4b21cf96d4cf612f239a6c322b10c8fe-Paper.pdf}.

\bibitem[Zakariaei et~al.(2024)Zakariaei, Rout, Haber, and Eliasof]{zakariaei2024advection}
Niloufar Zakariaei, Siddharth Rout, Eldad Haber, and Moshe Eliasof.
\newblock Advection augmented convolutional neural networks.
\newblock \emph{arXiv preprint arXiv:2406.19253}, 2024.

\bibitem[Lensink et~al.(2022)Lensink, Peters, and Haber]{lensink2019fully}
Keegan Lensink, Bas Peters, and Eldad Haber.
\newblock Fully hyperbolic convolutional neural networks.
\newblock \emph{Research in the Mathematical Sciences}, 9\penalty0 (4):\penalty0 1--22, 2022.

\bibitem[Kingma and Ba(2014)]{Kingma2014}
Diederik~P. Kingma and Jimmy Ba.
\newblock Adam: A method for stochastic optimization.
\newblock \emph{arXiv preprint arXiv:1412.6980}, 12 2014.
\newblock URL \url{http://arxiv.org/abs/1412.6980}.

\bibitem[Müller et~al.(2022)Müller, Schüler, Zech, and Heße]{Muller2022}
Sebastian Müller, Lennart Schüler, Alraune Zech, and Falk Heße.
\newblock Gstools v1.3: a toolbox for geostatistical modelling in python.
\newblock \emph{Geoscientific Model Development}, 15:\penalty0 3161--3182, 2022.

\bibitem[Knight et~al.(2018)Knight, Smith, Asch, Abraham, Cannia, Viezzoli, and Fogg]{knight2018mapping}
Rosemary Knight, Ryan Smith, Ted Asch, Jared Abraham, Jim Cannia, Andrea Viezzoli, and Graham Fogg.
\newblock Mapping aquifer systems with airborne electromagnetics in the central valley of california.
\newblock \emph{Groundwater}, 56\penalty0 (6):\penalty0 893--908, 2018.

\bibitem[West(2007)]{west2007water}
Fugro West.
\newblock Water resources investigation of the kaweah delta water conservation district.
\newblock \emph{Technical report, Project no. 3087.004. 07}, 2007.

\end{thebibliography}


\end{document}